# A multislice computational model for birefringent scattering

**SHUQI MU,**[1,2] **YINGTONG SHI,**[1] **YINTONG SONG,**[1] **WEI LIU,**[3] **WANXUE WEI,**[1,4] **QIHUANG GONG,**[1,2] **DASHAN DONG,**[1,4,*] **AND KEBIN SHI**[1,2,5]

[1]*State Key Laboratory for Mesoscopic Physics and Frontiers Science Center for Nano-optoelectronics, School of Physics, Peking University, Beijing 100871, China*
[2]*Collaborative Innovation Center of Extreme Optics, Shanxi University, Taiyuan 030006, China*
[3]*Aerospace Information Research Institute, Chinese Academy of Sciences, Beijing, 100094, China*
[4]*National Biomedical Imaging Center, Peking University, Beijing 100871, China*
[5]*Peking University Yangtze Delta Institute of Optoelectronics, Nantong 226010, China*
*\*dongdashan@pku.edu.cn*

**Abstract:** Modeling optical field propagation in highly scattering and birefringent medium is of important interest to many photonic research branches. Despite the existence of numerical electromagnetic simulation tools and beam propagation method frameworks, there has been a lack of an analytical model including the full tensor nature of birefringence, which is an essential forward-propagation tool for applications requiring efficiently iterative regularization and end-to-end designs. Here, we present an analytical tool for modeling field propagation in a birefringent scattering medium by including a full set of field tensor elements and multiple-scattering characteristics. Birefringence-controlled field propagation experiments were successfully carried out to validate the proposed model.

## 1. Introduction

The polarization state of photons undergoes a nonnegligible change as an optical wave propagates in a highly scattering medium, which can be modeled as a volume containing small randomly oriented birefringent compartments [1-3]. The intrinsic information about the scattered sample can be evaluated by measuring the depolarization degree, retardance and other vectorial properties [4] of the input polarized light or by directly detecting the appearance of the new polarization component in scattered vectorial light. Modeling this birefringent scattering is essential and can provide a useful tool to quantitatively analyze the interaction between a complex birefringent sample and incident polarized photons, which is urgently needed in the fields of studying the optical properties of semiconducting materials [5,6], designing polarization-related integrated devices [7-10], fabricating micro/nanocomponents [11,12], and imaging biomedical samples for clinical applications [13,14]. The commonly used numerical finite-difference time-domain (FDTD) method [15,16] and Monte Carlo method [17] can handle the propagation for electromagnetic problems of arbitrary computation sizes. However, the accuracy of the computation is sensitive to the sampling density of the computation volume, resulting in a requirement of high-speed access storage for large-scale simulation [15]. Moreover, the irreversible nature of these nonanalytical models makes solving the physically related inverse problems difficult, especially in the iterative reconstruction of three-dimensional (3D) learning-based computational imaging [18,19] and end-to-end optimization of optical devices [7]. Therefore, development of an efficient and reversible computation model to deal with complex birefringent scattering is necessary.

The scattering process brought by a birefringent sample is more complicated than the common scalar scattering [20-25], in which two factors need to be considered: the first is the natural characteristics of the birefringent objects, which will exhibit different responses according to the external polarization state and will induce polarization changes for the incident polarized light field [26]; the second is depolarization of the far-field propagation of the polarized light field [27]. Several studies on the vectorial beam propagation method (VBPM) have been reported in the past few years. The VBPM is based on the vectorial Helmholtz equations, and the birefringent effects of the material and polarization properties of the electric field are all considered in this method. However, due to the complexity of solving the vectorial wave equation, the VBPM is not feasible for numerical computation. Some approximations are usually made in the VBPM, including a weak longitudinal polarization component [19], paraxiality [26], and small birefringence [28]; these approximations largely limit its accuracy and applicability. More importantly, for ill-posed inverse scattering problems, this inaccurate method will introduce unacceptable errors, making its use in the field of 3D computational imaging difficult [29-31].

Here, we present a multislice computation model for birefringent scattering. The proposed model considers the complete polarization components of the vectorial field and includes the full elements of the 3 × 3 scattering potential tensor without neglecting the longitudinal (i.e., z-polarized) component. We exploit the dyadic Green's function and define a new polarization transfer function tensor (PTFT) to describe the polarization changes and depolarization during the scattering process. We verify the validity of our birefringent scattering model by simulating the vectorial light passing through anisotropic objects and comparing the results with those of the FDTD method. Then, we use the established optical tweezers-assisted polarization optical microscope to detect the vectorial scattering light field induced by synthetic birefringent samples to confirm the theoretical proposal. In both simulations and experiments, the proposed computational model is proven to be capable of high accuracy and efficiency.

## 2. Theory

### 2.1 Birefringent scattering model

The proposed model includes the three orthogonal polarization components by writing the electric field in the form of a column vector:

$$\vec{u}(\vec{r}) = \left[ u_x(\vec{r}), u_y(\vec{r}), u_z(\vec{r}) \right]^T, \tag{1}$$

where $\vec{u}(\vec{r})$ is the vectorial field and $u_x(\vec{r})$, $u_y(\vec{r})$ and $u_z(\vec{r})$ indicate the independent orthogonal polarization components of $\vec{u}(\vec{r})$ in Cartesian coordinates $\vec{r} = (x, y, z)$. During the propagation process, the birefringent objects can be viewed as spatially discrete scattering sources that perturb the polarization of the incident light and output different polarization components. Hence, the point spread function of the birefringent scattering should contain all the possible couplings between the input and output polarization components. As a result, it should be a 3 × 3 tensor. The 3D free-space dyadic Green's function [32] can fully describe the polarization changes after scattering by a birefringent sample, which has the following form:

$$\bar{\bar{G}}(\vec{r} - \vec{r}') = \left[ \bar{\bar{I}} + \frac{1}{k_m^2} \nabla \nabla \right] \frac{\exp(ik_m |\vec{r} - \vec{r}'|)}{4\pi |\vec{r} - \vec{r}'|}, \tag{2}$$

where $\vec{r} = (x, y, z)$ is the location of the evaluated field, where the coordinates $\vec{r}' = (x', y', z')$ designate the spatial position of the point source, $k_m = 2\pi n_m / \lambda$ denotes the wavenumber in the background medium, $n_m$ is the refractive index (RI) of the isotropic background medium, $\lambda$ is the incident light wavelength, and $\bar{\bar{I}}$ is the unit dyad. Compared with the scalar Green's function [33], the dyadic Green's function has a tensor part that relates all the polarization components of the source with all the polarization components of the scattered field.

The polarization coupling problem during the scattering process is complicated and highly nonlinear. The first-Born approximation is used here, and it only considers the single scattering of incident photons. This is valid when the scattered electric field is much smaller than the incident field. The total vectorial electric field $\vec{u}(\vec{r})$ is composed of an unscattered incident vectorial field $\vec{u}_{in}(\vec{r})$ and a singly scattered vectorial field $\vec{u}_s(\vec{r})$, where $\vec{u}_s(\vec{r})$ can be linearly related to the defined scattering potential tensor $\bar{\bar{V}}(\vec{r})$ by the dyadic Green's function. The total field can be formulated by the 3D dyadic Green's function-related first-order Born approximation [34], which is written as:

$$\vec{u}(\vec{r}) = \vec{u}_{in}(\vec{r}) + \iiint \bar{\bar{G}}(\vec{r}-\vec{r}') \times \bar{\bar{V}}(\vec{r}') \times \vec{u}_{in}(\vec{r}') d^3\vec{r}', \tag{3}$$

where $\bar{\bar{V}}(\vec{r})$ is the 3D scattering potential tensor of the object defined via the following expression:

$$\bar{\bar{V}}(\vec{r}) = k_0^2 \left( n_m^2 \bar{\bar{I}} - \bar{\bar{n}}(\vec{r})^2 \right), \tag{4}$$

with:

$$\bar{\bar{n}}(\vec{r}) = \begin{bmatrix} n_{xx}(\vec{r}) & n_{xy}(\vec{r}) & n_{xz}(\vec{r}) \\ n_{yx}(\vec{r}) & n_{yy}(\vec{r}) & n_{yz}(\vec{r}) \\ n_{zx}(\vec{r}) & n_{zy}(\vec{r}) & n_{zz}(\vec{r}) \end{bmatrix}, \tag{5}$$

where $k_0 = 2\pi/\lambda$ denotes the free-space wavenumber, and the symmetrical matrix $\bar{\bar{n}}(\vec{r})$ represents the RI distribution of the anisotropic birefringent sample at position $\vec{r} = (x, y, z)$. By calculating the spatial 3D deconvolution of Eq. (3), the total vectorial field after the light passes through the birefringent sample can be evaluated. This scattering model is essentially based on the vectorial nature of the polarization field without neglecting any longitudinal polarization component. The polarization coupling problem between the longitudinal and transverse polarization components imposed by the birefringent object is well described by using the dyadic Green's function.

*2.2 Polarization transfer function tensor*

Since the birefringent scattering model [Eq. (3)] describes the sample-induced polarization changes by the convolution relationship between the 3D point spread function tensor and scattering potential tensor, we can give a comprehensive description of the polarization coupling in the frequency domain by using the Fourier representation of the 3D dyadic Green's function. After imposing the constraints on the propagation vector and using Cauchy's residual theorem for Fourier integration, we obtain the forward 3D dyadic Green's function in terms of a continuous angular plane wave:

$$\bar{\bar{G}}(\vec{r}-\vec{r}') = \frac{i}{8\pi^2} \iint_{-\infty}^{+\infty} Q(k_x, k_y, k_z) \frac{\exp\{ik_x(x-x') + ik_y(y-y') + ik_z(z-z')\}}{k_z} dk_x dk_y, \tag{6}$$

with:

$$Q(k_x, k_y, k_z) = \begin{bmatrix} 1 - \frac{k_x^2}{k_m^2} & -\frac{k_x k_y}{k_m^2} & -\frac{k_x k_z}{k_m^2} \\ -\frac{k_y k_x}{k_m^2} & 1 - \frac{k_y^2}{k_m^2} & -\frac{k_y k_z}{k_m^2} \\ -\frac{k_z k_x}{k_m^2} & -\frac{k_z k_y}{k_m^2} & 1 - \frac{k_z^2}{k_m^2} \end{bmatrix}, \tag{7}$$

where, $k_x$, $k_y$ and $k_z$ are the spatial frequencies, which have the relationship $k_x^2 + k_y^2 + k_z^2 = k_m^2$. In Eq. (7), the square bracket $3 \times 3$ tensor part is what we call the polarization transfer function tensor (PTFT), represented by $Q(k_x, k_y, k_z)$. The elements in the tensor matrix $Q_{ij\{i,j=x,y,z\}}$ represent the coupling from the $i$ polarized source to the $j$ polarized scattered field in spatial frequencies $\vec{k} = (k_x, k_y, k_z)$. The amplitude and phase distribution of $Q(k_x, k_y, k_z)$ are shown in Fig. 1(a) and (b), respectively, and it is a symmetrical tensor matrix that corresponds to the 3D scattering potential tensor of the symmetrical object. A detailed derivation of the Fourier representation of the 3D dyadic Green's function can be seen in Section 1 of Supplement 1.

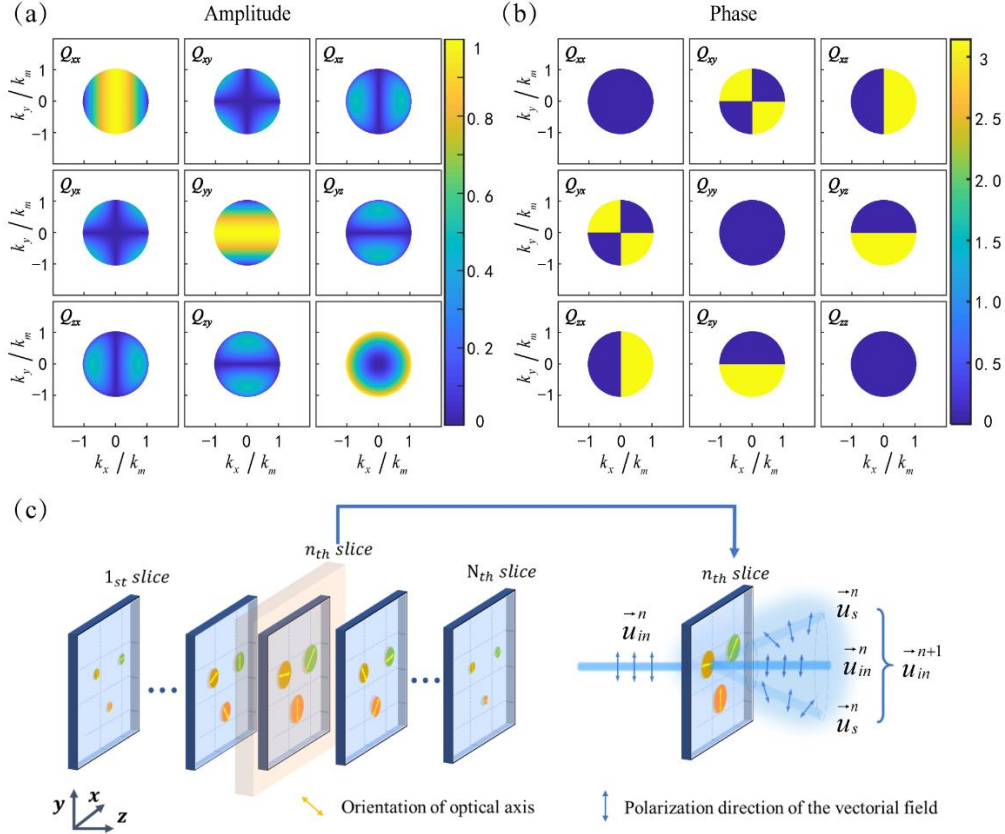

Fig. 1. Illustration of our proposed multislice computational model for birefringent scattering. (a), (b) Amplitude and phase distribution of the PTFT. (c) The three-dimensional birefringent samples are decomposed into multiple thin slices. The total vectorial electric field is composed of an unscattered field and a singly scattered field for every scattering slice, and its polarization will change during the process.

### *2.3 Multislice birefringent scattering algorithm*

In the scattering model [Eq. (3)], the scattering potential only interacts with the unscattered vectorial incident field, and the multiple scattering during the propagation is neglected; this approximation is only suitable for a thin sample. To overcome this restriction and make the model computationally accurate, we use the previously reported multislice method [22] to improve the applicability and efficiency of the birefringent scattering model. We divide the 3D scattering potential into multiple 2D slices with finite thickness Δz along the axial direction, as shown in Fig. 1(c). For each slice, we apply the 3D dyadic Green's function-related first-order Born approximation, and we can then obtain the vectorial multislice model suitable for birefringent scattering. The forward propagation model can be written as:

$$\vec{u}_{in}^{n+1}\left[x,y,(n+1)\Delta z\right]=\vec{u}_{in}^{n}\left[x,y,(n+1)\Delta z\right]+\vec{u}_{s}^{n}\left[x,y,(n+1)\Delta z\right], \tag{8}$$

and:

$$\vec{u}_{s}^{n}\left[x,y,(n+1)\Delta z\right]=$$
$$-\int_{0}^{\Delta z}\iint \overline{\overline{G}}(x-x',y-y',\Delta z-\varepsilon)\overline{\overline{V}}(x',y',n\Delta z+\varepsilon)\vec{u}_{in}^{n}\left[x',y',n\Delta z+\varepsilon\right]dx'dy'd\varepsilon \tag{9}$$

where $n$ indicates the $n_{th}$ slice with total axial discrete number $N$. In Eq. (8), the $(n+1)_{th}$ incident electric field $\vec{u}_{in}^{n+1}[x,y,(n+1)\Delta z]$ is composed of two components: the first part is the $n_{th}$ incident light $\vec{u}_{in}^{n}[x,y,n\Delta z]$ that propagates one slice forward, and the second part is the scattering light $\vec{u}_{s}^{n}[x,y,(n+1)\Delta z]$ induced by the anisotropic sample of the $n_{th}$ slice. By recursively using Eqs. (8) and (9) for each slice, the total vectorial field passing through the birefringent sample can be accurately calculated.

The vectorial free-space propagation to the $(n+1)_{th}$ slice must first be calculated based on the given boundary field, which is the incident vectorial field $\vec{u}_{in}^{n}(x,y,n\Delta z)$ at slice $n_{th}$. The solution of this problem belongs to the classical Dirichlet problem [35], and it can be approached by applying the method of images [36] to the vectorial field. In this case, a modified 3D dyadic Green's function for a single scattering slice has the form:

$$\overline{\overline{G}}(\vec{r},\vec{r}')_{image}=\left[\overline{\overline{I}}+\frac{1}{k_m^2}\nabla\nabla\right]\left(\frac{\exp(ik_m R)}{4\pi R}-\frac{\exp(ik_m R_1)}{4\pi R_1}\right) \tag{10}$$

with:

$$R=\sqrt{(x-x')^2+(y-y')^2+(z-z')^2},$$
$$R_1=\sqrt{(x-x')^2+(y-y')^2+(z+z')^2},$$

The modified 3D dyadic Green's function subtracts image term $R_1$ to satisfy the boundary condition. By using Eq. (10) and considering the radiation condition in the far field, the vectorial free-space propagation process can be formulated as:

$$\vec{u}_{in}^{n}(\vec{r})=-\iint_{z'=n\Delta z}\vec{u}_{in}^{n}(\vec{r}')\frac{\partial \overline{\overline{G}}(\vec{r},\vec{r}')_{image}}{\partial \vec{n}}dx'dy' \tag{11}$$

Let $\vec{r}=(x,y,(n+1)\Delta z)$, $\vec{r}'=(x',y',n\Delta z)$, and normal vector $\vec{n}$ be on the boundary points in the $-z$ direction. Substituting Eqs. (6), (7) and (10) into (11), the vectorial free-space propagation process can be written as:

$$\vec{u}_{in}^{n}(x,y,(n+1)\Delta z)=\mathbf{P}\,\vec{u}_{in}^{n}(x,y,n\Delta z) \tag{12}$$

where the tensor matrix $\mathbf{P}$ is the vectorial free-space propagation operator:

$$\mathbf{P}=\mathcal{F}^{-1}Q(k_x,k_y,k_z)\exp(ik_z\Delta z)\mathcal{F}\{\cdots\}, \tag{13}$$

where $\mathcal{F}$ and $\mathcal{F}^{-1}$ are the 2D Fourier transform and inverse Fourier transform, respectively, which can be performed by the numerical fast Fourier transform (FFT). The operator $\mathbf{P}$ can describe the depolarization of the far-field propagation of the polarized light field, and it is similar to the scalar angular spectrum propagation kernel, which simply drops the PTFT part in Eq. (13). See Section 2 of Supplement 1 for a detailed derivation of the vectorial free-space propagation process.

Meanwhile, based on the convolution theorem, assuming that the scattering potential tensor does not vary axially within each slice, we can simplify the forward birefringent scattering Eq. (9) as:

$$\vec{u}_s^n[x,y,(n+1)\Delta z] = \mathbf{H}\,\overline{\overline{V}}(x,y,n\Delta z)\mathbf{Q}\,\vec{u}_{in}^n[x,y,n\Delta z], \tag{14}$$

where the tensor matrix $\mathbf{Q}$ is the polarization transfer operator:

$$\mathbf{Q} = \mathcal{F}^{-1} Q(k_x,k_y,k_z)\mathcal{F}\{\cdots\} \tag{15}$$

and the tensor matrix $\mathbf{H}$ denotes the vectorial scattering operator:

$$\mathbf{H} = \mathcal{F}^{-1} -\frac{i}{2}Q(k_x,k_y,k_z)\frac{\exp(ik_z\Delta z)}{k_z}\mathcal{F}\{\cdots\}\Delta z, \tag{16}$$

The vectorial scattering operator $\mathbf{H}$ describes the sample scattering-induced polarization change. By sequentially using Eq. (11) and Eq. (14) on each slice, the 3D forward vectorial field can be efficiently evaluated. See Section 3 of Supplement 1 for the derivation of the forward birefringent scattering process.

## 3. Simulation verification

We confirmed the accuracy of the proposed computational model by simulating the vectorial light passing through a birefringent object and comparing the scattering results with those of the FDTD method. In our simulation, a set of birefringent digital phantoms with known birefringent tensors were chosen to demonstrate the verification. As shown in Fig. 2(a), four 3 $\mu m$-diameter birefringent beads were placed in a homogenous background medium with RI $n_m = 1.333$. We set the principle RIs as $n_{xx} = 1.37$, $n_{yy} = 1.40$ and $n_{zz} = 1.44$. For the arbitrary optical axis orientation of the birefringent sample, we can define the RI tensor by applying a 3D rotational transform matrix $R$ to the principle dielectric tensor.

$$\mathrm{RI}^2 = R(\theta_x,\theta_y,\theta_z)\begin{pmatrix} n_{xx}^2 & 0 & 0 \\ 0 & n_{yy}^2 & 0 \\ 0 & 0 & n_{zz}^2 \end{pmatrix} R(\theta_x,\theta_y,\theta_z)^T, \tag{17}$$

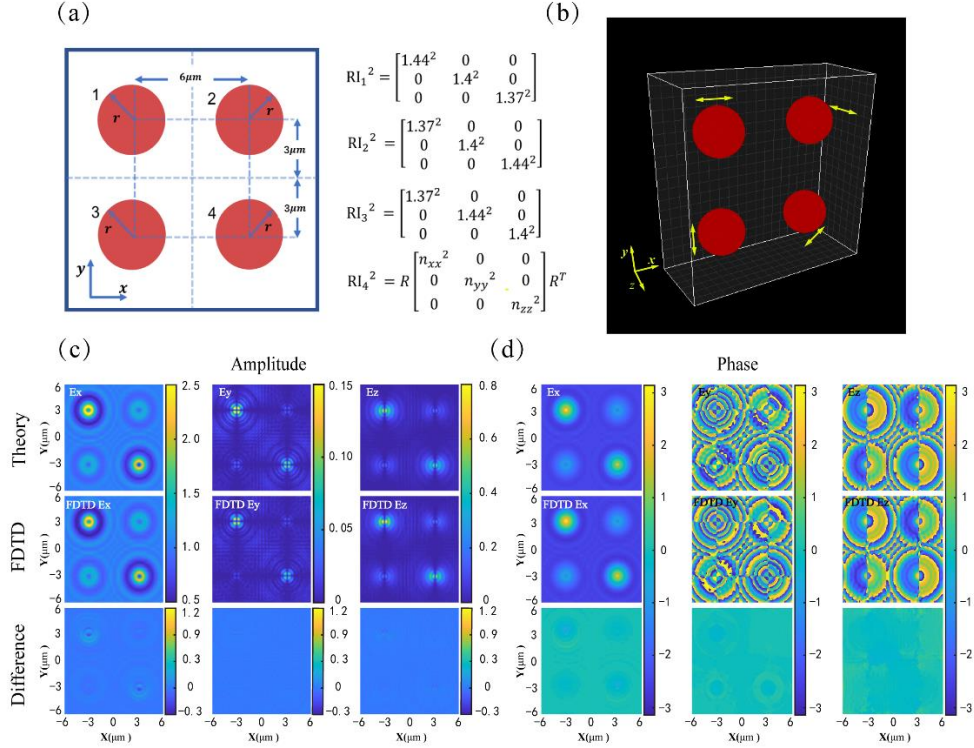

Fig. 2. Simulation of the birefringent scattering by our proposed model and the FDTD method. (a) Parameter used to define the birefringent spheres. (b) 3D rendering of the birefringent spheres. The double yellow arrows indicate the birefringence orientations. (c), (d) Amplitude and phase distribution of three orthogonal polarization components $E_x$, $E_y$ and $E_z$ after the light passes through the birefringent samples. The first row shows the theoretical results given by our proposed model, the second row shows the simulated results from the FDTD method, and the final row shows the discrepancies between the two methods.

where $\theta_x$, $\theta_y$ and $\theta_z$ are the rotation angles around the $x$, $y$, and $z$ axes, respectively. To study the effect of the birefringence orientation on the incident polarized light, we made the four birefringent beads have different birefringence orientations, and the RI tensors for the four spheres are presented in Fig. 2(a). The optical axes of the first three spheres were set parallel to the coordinate axes but mutually orthogonal, and that of the fourth sphere was set to $\theta_x = \theta_y = \theta_z = \frac{\pi}{4}$. Fig. 2(b) shows the 3D organization of the four spheres, where the double yellow arrows indicate the birefringence orientations. The total computational volume was set to $11.7\ \mu m \times 11.7\ \mu m \times 4.55\ \mu m$ with a sampling size of 65 nm in all three directions. In our simulation, we used an x-polarized 405 nm plane wave with an equal amplitude of 1 as the incident light. Fig. 2(c) and (d) shows the amplitude and phase distribution of the three orthogonal polarization components $E_x$, $E_y$ and $E_z$ after the light passes through the birefringent samples. There are still some differences between the two methods. The FDTD method considers both forward and backward scattering in the simulation, while the proposed model only calculates the forward information. However, these discrepancies are small and can be neglected in both amplitude and phase, as shown in Fig. 2(c) and (d), which shows that our model can achieve the same accurate results as the FDTD method. Importantly, running on a workstation equipped with 2× Inter Xeon E5-2697A V4, 2.6 GHz 128 GB, the computation time of our proposed model was only 0.77 s, which is significantly faster than that of the FDTD method (35 s). From the simulation, we can clearly identify that different orientations of the optical axis of the sample will cause various polarization coupling values, which is more obvious for sphere 1, whose $n_{xx}$ is 1.44, resulting in the strongest coupling compared with the other spheres. Moreover, the longitudinal polarization component $E_z$ has intensity comparable to those of the transverse components $E_x$ and $E_y$. Hence, ignoring the axial polarization component for these highly birefringent scattering objects is no longer appropriate.

### 4. Experimental verification

To verify the proposed model, we built an optical tweezer-assisted polarization optical microscope to measure the vectorial scattering light field of birefringent samples. A schematic diagram of the experiment is shown in Fig. 3(a). For the imaging part, to reduce the coherent and environmental noise, we used a fiber-coupled LED (M405FP1, Thorlabs) as the imaging source. The x-polarized collimated imaging light first passed through a 4f system comprising an achromatic lens L1 (AC254-100-A, Thorlabs) and a water-immersion objective (Obj1, 60×, NA=1.0, LUMPLFLN60XW). The light passing through the birefringent sample was collected using another water-immersion objective (Obj2, 60×, NA=1.2, UPLSAPO60XW, Olympus) and collimated by L2 (ACT508-400-A, Thorlabs). A polarizer P (LPUV100, Thorlabs) was placed in front of a sCMOS camera (ORCA-Flash 4.0 V3, Hamamastu), and it was rotated to filter the corresponding polarization component of the scattering light. For the trapping part, a single longitudinal model laser ($\lambda = 473$nm, 200 mW, CNI) was used as the trapping beam. After collimation by lens L3 (AC254-040-A, Thorlabs), the polarized trapping light passed through a half-wave plate HWP (WPH10-405, Thorlabs), which could control the polarization of the light. The beam was then reflected by a pair of mirrors to the back focal plane of the objective lens and formed optical trapping after being focused by Obj2. The imaging light and trapping light were separated by a long-pass dichroic mirror DM (DMLP425R, Thorlabs).

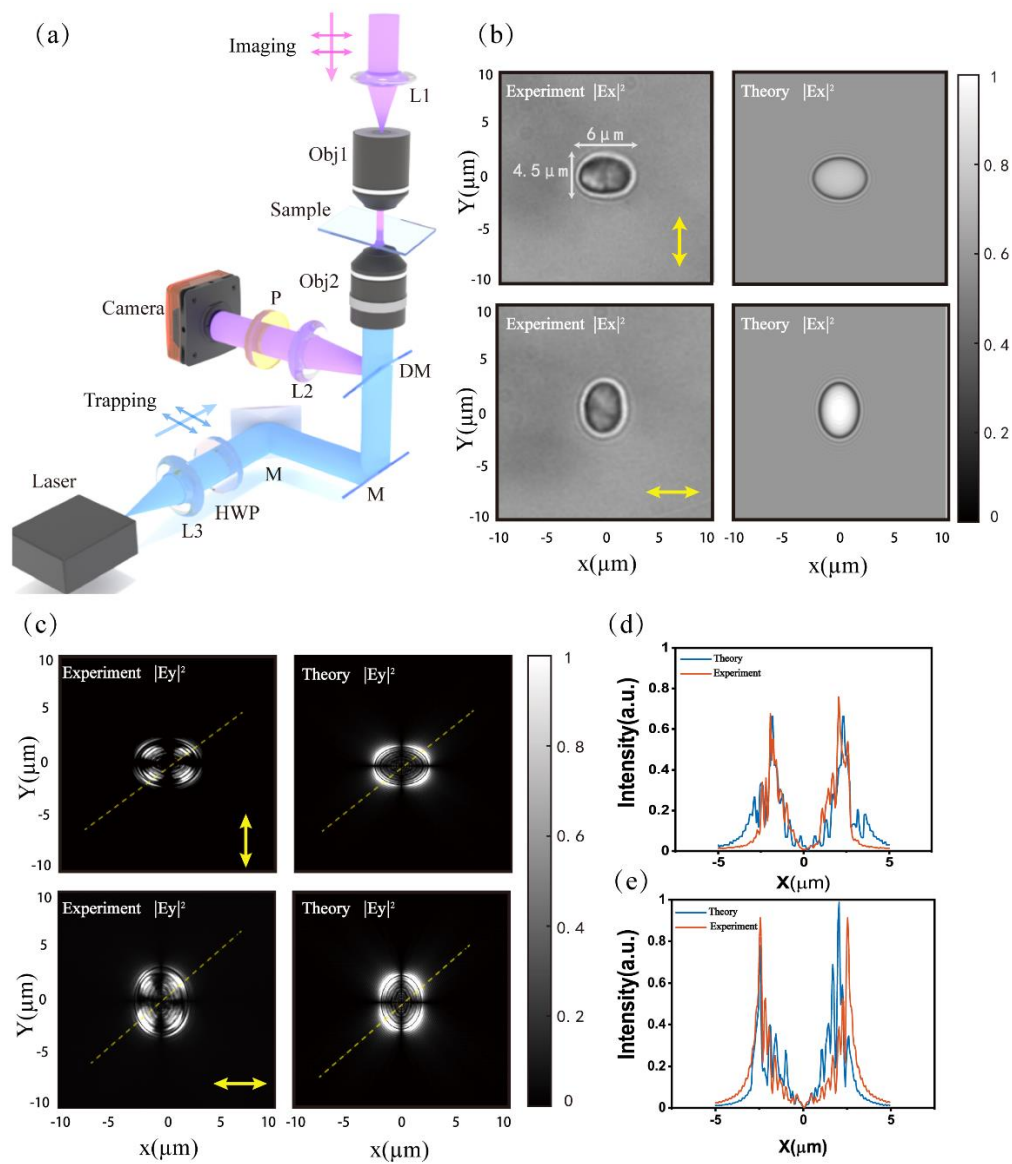

Fig. 3. Optical tweezers-assisted polarization optical microscope to detect the vectorial scattering light field. (a) Experimental setup. (b) Experimental and simulation results of the x-polarized scattering light $|E_x|^2$ for two different optical axis orientations of the vaterite particles. (c) Experimental and simulation results of the cross-polarized scattering light $|E_y|^2$ for two different optical axis orientations of the vaterite particles. The yellow double axes in (b) and (c) indicate the slow axis of vaterite. (d), (e) Intensity distribution taken from the yellow dashed line in (c).

The birefringent samples used in our experiment were vaterite particles 4-6 μm in diameter. The vaterite particles have a polycarbonate structure of calcium carbonate composed of 20-30 nm size nanocrystals with positive uniaxial birefringence [37-39]. The alignment of the optical axes of the vaterite particles has a hyperbolic distribution throughout the volume, which makes vaterite have a high birefringence coefficient of up to 0.1 [40,41]. Using the half-wave plate

HWP in the trapping beam setup, the orientation of the trapped vaterite particles can be precisely rotated to any angle about the beam axis.

The size of the vaterite particles is essential for detecting the vectorial scattering light. Particles with small sizes cannot cause sufficient scattering and will bring very large discrepancies upon comparison with the theoretical results. Hence, we modified the synthesis of vaterite particles based on a previously published protocol [37,38]. Aqueous solutions of $CaCl_2$, $MgSO_4$, and $K_2CO_3$ were prepared with a molarity of 0.1M . First, 1.5 mL of $CaCl_2$ was mixed with 60 μL of $MgSO_4$ in a 5 mL reaction vessel, followed by 90 μL of $K_2CO_3$. The solution was agitated by violently pipetting the solution for 4 minutes. After that, 100 μL of a surfactant solution (XYS-3500, Yancheng Yunfeng Chemical Co., Ltd.) was added to the solution to stabilize the reaction. The concentration of the surfactant solution is essential to the growth of vaterite particles. We dissolved 2.2 mg of the surfactant in 200 mL of distilled water and achieved the expected particle size.

The experimental and simulation results are presented in Fig. 3. Using the trapping beam, we can orient the optical axis of the vaterite particles parallel or perpendicular to the polarization of the incident imaging light and detect the transverse polarization components of the scattering light. A trapped particle is shown in Fig. 3(b) and (c), where the yellow double axes indicate the optical axis (slow axis) of vaterite. The experimental results obviously show good agreement with the corresponding theoretical prediction, especially in the orthogonal measurement (Fig. 3(c)), where the polarization axis of polarizer P is perpendicular to the polarization of the imaging light (x polarized), and the high extinction ratio gives a clearer cross-polarized $|E_y|^2$ pattern. When the optical axis of the birefringent vaterite particle is parallel to the polarization of the imaging light, the polarization coupling will be significantly enhanced, and we plotted the distribution of the normalized $|E_y|^2$ field, as shown in Fig. 3(d) and (e). The experimental results further validate the accuracy of our scattering model. Note that the synthesized vaterite particles are not perfectly homogenous in RI, which leads to disagreement between experiments and simulations in the $|E_x|^2$ field, as shown in Fig. 3(b).

## 5. Discussion and conclusions

In conclusion, we propose a multislice computational model for birefringent scattering. Taking the complete vectorial polarization components into consideration, our proposed model is unambiguous by using the dyadic Green's function. We then define a new PTFT to describe the polarization changes and depolarization during the scattering process. The vectorial properties of the electric field associated with the birefringent material are fully considered without neglecting any longitudinal polarization component in our model. The theoretical and experimental results clearly support the general applicability of this computational model and prove that it is capable of performing various scattering calculations with high accuracy and efficiency. With these unique advantages, we believe that this model will provide a solution to tackle the challenges of large-scale 3D reconstruction/design of computational imaging and photonic devices.

**Funding.** This work was funded by Guangdong Major Project of Basic and Applied Basic Research No. 2020B0301030009, the National Natural Science Foundation of China (92150301, 91750203, 12041602, 91850111, 12004012, and 12004013), China Postdoctoral Science Foundation (2020M680230, 2020M680220), Clinical Medicine Plus X - Young Scholars Project, Peking University, the Fundamental Research Funds for the Central Universities, and the High-performance Computing Platform of Peking University.

**Acknowledgments.**

**Disclosures.** We declare that we have no conflicts of interest.

**Data availability.** All data included in this study are available upon request by contacting the corresponding author.

**Supplemental document.** See Supplement 1 for supporting content.

# A multislice computational model for birefringent scattering: supplemental document

This document provides supplementary information to "A multislice computational model for birefringent scattering". Here, we include all the derivations of our proposed computational model. We first show the Fourier representation of the 3D dyadic Green's function and define the polarization transfer function tensor (PTFT) to describe the polarization change during the scattering process. Then, we give the derivation of the vectorial free-space propagation and forward birefringent scattering based on the PTFT.

## 1. Derivation of the fourier representation of the 3D dyadic Green's function

In this section, we give a detailed derivation process for the Fourier transform of the 3D dyadic Green's function [1]. The 3D dyadic Green's function can be written as:

$$\overline{\overline{G}}(\vec{r}-\vec{r}') = \left[\overline{\overline{I}} + \frac{1}{k_m^2}\nabla\nabla\right]g(\vec{r}-\vec{r}'), \tag{S1}$$

where $g(\vec{r}-\vec{r}')$ is the scalar Green's function. In Cartesian coordinates, the gradient operator $\nabla$ can be written as:

$$\nabla = \vec{e}_x \frac{\partial}{\partial x} + \vec{e}_y \frac{\partial}{\partial y} + \vec{e}_z \frac{\partial}{\partial z} \tag{S2}$$

where $\vec{e}_x$, $\vec{e}_y$ and $\vec{e}_z$ are the unit vectors in the $x$, $y$ and $z$ directions, respectively. Then, the 3D dyadic Green's function can be described by:

$$\overline{\overline{G}}(\vec{r}-\vec{r}') = \begin{bmatrix} k_m^2 + \frac{\partial^2}{\partial x^2} & \frac{\partial^2}{\partial x \partial y} & \frac{\partial^2}{\partial x \partial z} \\ \frac{\partial^2}{\partial y \partial x} & k_m^2 + \frac{\partial^2}{\partial y^2} & \frac{\partial^2}{\partial y \partial z} \\ \frac{\partial^2}{\partial z \partial x} & \frac{\partial^2}{\partial z \partial y} & k_m^2 + \frac{\partial^2}{\partial z^2} \end{bmatrix} \frac{1}{k_m^2} g(\vec{r}-\vec{r}'), \tag{S3}$$

We start from the scalar Green's function. To simplify the derivation process, without loss of generality, we assume that the location of the source point is at the origin. The scalar Green's function $g(\vec{r})$ satisfies the scalar wave equation:

$$\left(\frac{\partial^2}{\partial x^2} + \frac{\partial^2}{\partial y^2} + \frac{\partial^2}{\partial z^2}\right)g(\vec{r}) = -\delta(\vec{r}), \tag{S4}$$

The Fourier transform of $g(\vec{r})$ is represented by $\tilde{g}(\vec{k})$, and the relationship between $g(\vec{r})$ and $\tilde{g}(\vec{k})$ is given by:

$$\tilde{g}(\vec{k}) = \iiint_{-\infty}^{+\infty} g(\vec{r}) e^{-i(k_x x + k_y y + k_z z)} dx dy dz, \tag{S5}$$

and:

$$g(\vec{r}) = \frac{1}{(2\pi)^3} \iiint_{-\infty}^{+\infty} \tilde{g}(\vec{k}) e^{i(k_x x + k_y y + k_z z)} dk_x dk_y dk_z, \tag{S6}$$

where the spatial frequencies $k_x$, $k_y$ and $k_z$ have the relationship $k_x^2 + k_y^2 + k_z^2 = k_m^2$. Substituting Eq. (S6) into Eq. (S5) and noting that the 3D Dirac function $\delta(\vec{r})$ has the property:

$$\delta(\vec{r}) = \frac{1}{(2\pi)^3} \iiint_{-\infty}^{+\infty} e^{i(k_x x + k_y y + k_z z)} dk_x dk_y dk_z, \tag{S7}$$

we obtain $\tilde{g}(\vec{k})$:

$$\tilde{g}(\vec{k}) = \frac{1}{k_x^2 + k_y^2 + k_z^2 - k_m^2}, \tag{S8}$$

Substituting Eq. (S8) into Eq. (S6), we obtain:

$$g(\vec{r}) = \frac{1}{(2\pi)^3} \iiint_{-\infty}^{+\infty} \frac{1}{k_x^2 + k_y^2 + k_z^2 - k_m^2} e^{i(k_x x + k_y y + k_z z)} dk_x dk_y dk_z, \tag{S9}$$

Eq. (S8) gives the Fourier transform of the scalar Green's function; however, the three spatial components are necessarily independent. Commonly speaking, knowing the transverse frequency, the corresponding axial frequency can be directly calculated by $k_z^2 = k_m^2 - k_x^2 - k_y^2$. Therefore, the integration constraint with respect to $k_z$ is added by using the auxiliary variable $h$, and we obtain the integration:

$$g(\vec{r}) = \frac{1}{(2\pi)^2} \iint_{-\infty}^{+\infty} h(z) e^{i(k_x x + k_y y)} dk_x dk_y, \tag{S10}$$

with:

$$h(z) = \frac{1}{2\pi} \int_{-\infty}^{+\infty} \frac{1}{k_z^2 - h^2} e^{ik_z z} dk_z, \quad \text{with} \quad h^2 = k_m^2 - k_x^2 - k_y^2, \tag{S11}$$

Eq. (S11) can be calculated by the residue theorem as:

$$h(z) = \frac{i}{2h} e^{ih|z|}, \quad \text{with} \quad h^2 = k_m^2 - k_x^2 - k_y^2, \tag{S12}$$

Eq. (S12) gives the expression for the integration with respect to $k_z$. $h(z)$ is a piecewise function; the $z \geq 0$ part describes the propagation along the $+z$ direction, and the $z < 0$ part describes the propagation along the $-z$ direction. For our transmission type of propagation model, we only consider the positive part. We replace $h$ with $k_z$ to constrain the integration $h(z)$, and we finally obtain:

$$h(z) = \frac{i}{2k_z} e^{ik_z z}, \tag{S13}$$

Substituting Eq. (S13) into Eq. (S10):

$$g(\vec{r}) = \frac{i}{8\pi^2} \iint_{-\infty}^{+\infty} \frac{e^{i(k_x x + k_y y + k_z z)}}{k_z} dk_x dk_y, \quad \text{with} \quad k_x^2 + k_y^2 + k_z^2 = k_m^2, \tag{S14}$$

Substituting Eq. (S14) into Eq. (S3), we note that the differential operators $\frac{\partial}{\partial x}, \frac{\partial}{\partial y}$ and $\frac{\partial}{\partial z}$ in Eq. (S3) can be replaced by $ik_x$, $ik_y$ and $ik_z$. The Fourier transform of the dyadic Green's function can be shown as:

$$\bar{\bar{G}}(\vec{r}-\vec{r}') = \frac{i}{8\pi^2} \int\int_{-\infty}^{+\infty} Q(k_x,k_y,k_z) \frac{e^{i(k_x x + k_y y + k_z z)}}{k_z} dk_x dk_y, \tag{S15}$$

where the square bracket 3 × 3 tensor part is what we define as the PTFT, and it is represented by $Q(k_x, k_y, k_z)$:

$$Q(k_x,k_y,k_z) = \begin{pmatrix} 1-\frac{k_x^2}{k_m^2} & -\frac{k_x k_y}{k_m^2} & -\frac{k_x k_z}{k_m^2} \\ -\frac{k_y k_x}{k_m^2} & 1-\frac{k_y^2}{k_m^2} & -\frac{k_y k_z}{k_m^2} \\ -\frac{k_z k_x}{k_m^2} & -\frac{k_z k_y}{k_m^2} & 1-\frac{k_z^2}{k_m^2} \end{pmatrix} = \begin{pmatrix} Q_{xx} & Q_{xy} & Q_{xz} \\ Q_{yx} & Q_{yy} & Q_{yz} \\ Q_{zx} & Q_{zy} & Q_{zz} \end{pmatrix}, \tag{S16}$$

The amplitude and phase distribution of the elements in the polarization optical function are shown in Fig. 1(a) and (b).

## 2. Derivation of the vectorial free-space propagation process

Poisson's equation is appropriate for describing the relationship between the source and electric field. In free-space propagation, the vectorial field propagates in a homogenous background medium without the scattering source, and the modified Poisson's equation can be expressed as:

$$\vec{u}(\vec{r}) = \iint_{\Sigma} \left[ \bar{\bar{G}}(\vec{r}-\vec{r}') \frac{\partial \vec{u}(\vec{r}')}{\partial \vec{n}} - \vec{u}(\vec{r}') \frac{\partial \bar{\bar{G}}(\vec{r}-\vec{r}')}{\partial \vec{n}} \right] d\vec{r}', \tag{S17}$$

where the integration is calculated on the closed surface $\Sigma$, and $\vec{n}$ is the normal vector on the surface $\Sigma$. We define the closed surface $\Sigma$ as being composed of the plane $z'=0$ and the hemispherical surface with infinite radius $\Sigma_{R\to\infty}$ of $z'>0$, as shown in Fig. S1. After considering the radiation condition where the integration Eq. (S17) is zero on $\Sigma_{R\to\infty}$, the simplified Poisson's equation can be written as:

$$\vec{u}(\vec{r}) = \iint_{z'=0} \left[ \bar{\bar{G}}(\vec{r}-\vec{r}') \frac{\partial \vec{u}(\vec{r}')}{\partial \vec{n}} - \vec{u}(\vec{r}') \frac{\partial \bar{\bar{G}}(\vec{r}-\vec{r}')}{\partial \vec{n}} \right] dx'dy', \tag{S18}$$

Without loss of generality, we assume that the input electric field is at the plane $z'=0$. This problem of vectorial free-space propagation belongs to the Dirichlet problem [2].

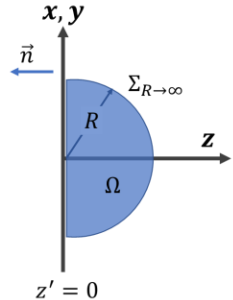

Fig. S1. Illustration of the vectorial free-space propagation.

The corresponding dyadic Green's function satisfies the equation:

$$\begin{cases} \nabla \times \nabla \times \overline{\overline{G}}(\vec{r}-\vec{r}') - k_m^2 \overline{\overline{G}}(\vec{r}-\vec{r}') = \overline{\overline{I}}\,\delta(\vec{r}-\vec{r}'), \\ \overline{\overline{G}}(\vec{r}-\vec{r}')\big|_{z'=0} = 0, \end{cases} \quad (S19)$$

Substituting Eq. (S19) into Eq. (S18), we can obtain:

$$\vec{u}(\vec{r}) = -\iint_{z'=0} \vec{u}(\vec{r}') \frac{\partial \overline{\overline{G}}(\vec{r}-\vec{r}')}{\partial \vec{n}} dx'dy', \quad (S20)$$

A new mirror-image dyadic Green's function following the boundary condition of Eq. (S19) can be derived by the method of images [3] and can be written as:

$$\overline{\overline{G}}(\vec{r},\vec{r}')_{image} = \left[\overline{\overline{I}} + \frac{1}{k_m^2}\nabla\nabla\right]\left(\frac{\exp(ik_m R)}{4\pi R} - \frac{\exp(ik_m R_1)}{4\pi R_1}\right), \quad (S21)$$

with:

$$R = \sqrt{(x-x')^2 + (y-y')^2 + (z-z')^2},$$
$$R_1 = \sqrt{(x-x')^2 + (y-y')^2 + (z+z')^2},$$

Compared with the free-space dyadic Green's function Eq. (S1), the imaged dyadic Green's function subtracts the image term to satisfy the boundary condition of the electric field at the plane $z'=0$. The normal vector $\vec{n}$ on plane $z'=0$ is in the negative direction of the z axis. Substituting Eq. (S21) and Eq. (S15) into Eq. (S18), we can obtain:

$$\vec{u}(\vec{r}) = \iint_{z'=0} \vec{u}(\vec{r}') \frac{1}{4\pi^2} \int\int_{-\infty}^{+\infty} Q(k_x,k_y,k_z) e^{ik_x(x-x')+ik_y(y-y')+ik_z(z-z')} dk_x dk_y dxdy,$$
$$(S22)$$

For Eq. (S22), we exchange the integration orders and perform the Fourier transform on both sides, and we can obtain the vectorial free-space propagation:

$$\vec{u}(k_x,k_y,z) = Q(k_x,k_y,k_z) e^{ik_z z} \vec{u}(k_x,k_y,0), \quad (S23)$$

where $\vec{u}(k_x,k_y,0)$ is the Fourier transform of the incident vectorial field $\vec{u}(x,y,0)$ and $\vec{u}(k_x,k_z,z)$ is the Fourier transform of the vectorial field $\vec{u}(x,y,z)$ that propagates from $\vec{u}(x,y,0)$ by distance z. We rewrite Eq. (S23) by using the vectorial free-space operator **P**:

$$\vec{u}_{in}^n(x,y,(n+1)\Delta z) = \mathbf{P}\,\vec{u}_{in}^n(x,y,n\Delta z), \quad (S24)$$

where the tensor matrix **P** is the defined vectorial free-space propagation operator:

$$\mathbf{P} = \mathcal{F}^{-1} Q(k_x,k_y,k_z) \exp(ik_z \Delta z) \mathcal{F}\{\cdots\}, \quad (S25)$$

where $\mathcal{F}$ and $\mathcal{F}^{-1}$ are the 2D Fourier transform and inverse Fourier transform, respectively. Compared with the scalar angular spectrum propagation method, which only has the angular spectrum $e^{ik_z z}$ term, Eq. (S23) can describe the far-field depolarization by using the PTFT.

### 3. Derivation of the forward birefringent scattering process

For the forward birefringent scattering process, the first-Born approximation is used on every slice [4], and the multiple vectorial scattering effects are considered in the propagation model.

$$\vec{u}_s^n[x,y,(n+1)\Delta z] = \\ -\int_0^{\Delta z}\iint \overline{\overline{G}}(x-x',y-y',\Delta z-\varepsilon)\overline{\overline{V}}(x',y',n\Delta z+\varepsilon)\vec{u}_{in}^n[x',y',n\Delta z+\varepsilon]dx'dy'd\varepsilon, \quad (S26)$$

Substituting Eq. (S15) into Eq. (S26), we can obtain:

$$\vec{u}_s^n[x,y,(n+1)\Delta z] = -\int_0^{\Delta z}\iint \frac{i}{8\pi^2}\iint Q(k_x,k_y,k_z)\frac{1}{k_z}e^{ik_x(x-x')+ik_y(y-y')+ik_z(\Delta z-\varepsilon)}dk_x dk_y \\ \cdot \overline{\overline{V}}(x',y',n\Delta z+\varepsilon)\vec{u}_{in}^n[x',y',n\Delta z+\varepsilon]dx'dy'd\varepsilon, \quad (S27)$$

Simplifying Eq. (S27):

$$\vec{u}_s^n[x,y,(n+1)\Delta z] = -\frac{i}{8\pi^2}\int_0^{\Delta z}\iint Q(k_x,k_y,k_z)\frac{1}{k_z}e^{i(k_x x+k_y y+k_z(\Delta z-\varepsilon))}dk_x dk_y \\ \cdot \iint \overline{\overline{V}}(x',y',n\Delta z+\varepsilon)\vec{u}_{in}^n[x',y',n\Delta z+\varepsilon]e^{-i(k_x x'+k_y y')}dx'dy'\, d\varepsilon, \quad (S28)$$

Performing the Fourier transform for the 2D coordinates $(x,y)$ on both sides of Eq. (S28) and noting that the 2D Dirac function $\delta(k_x,k_y)$ has the property:

$$\delta(k_x,k_y) = \frac{1}{2\pi^2}\iint e^{i(k_x x+k_y y+k_z z)}dxdy, \quad (S29)$$

we can finally obtain:

$$\vec{u}_s^n[k_x^D,k_y^D,(n+1)\Delta z] = -\frac{i}{2}\int_0^{\Delta z}Q(k_x^D,k_y^D,k_z^D)\frac{1}{k_z^D}e^{ik_z^D(\Delta z-\varepsilon)} \\ \cdot \iint \overline{\overline{V}}(x',y',n\Delta z+\varepsilon)\vec{u}_{in}^n[x',y',n\Delta z+\varepsilon]e^{-i(k_x^D x'+k_y^D y')}dx'dy'\, d\varepsilon, \quad (S30)$$

where $(k_x^D, k_y^D)$ is the 2D spatial frequency according to $(x,y)$. $\vec{u}_{in}^n[x',y',n\Delta z+\varepsilon]$ is the incident vectorial field $\vec{u}_{in}^n[x',y',n\Delta z]$ that propagates by distance $\varepsilon$; thus, it can be calculated by our proposed vectorial free-space propagation method. Substituting Eq. (S23) into Eq. (S30), we can obtain:

$$\vec{u}_s^n[k_x^D,k_y^D,(n+1)\Delta z] = -\frac{i}{2}Q(k_x^D,k_y^D,k_z^D)\frac{1}{k_z^D}e^{ik_z^D \Delta z} \\ \cdot \iint \overline{\overline{V}}(k_x^D-k_x^{D'},k_y^D-k_y^{D'},n\Delta z)Q(k_x^{D'},k_y^{D'},k_z^{D'})\vec{u}_{in}^n[k_x^{D'},k_y^{D'},n\Delta z]dk_x^{D'}k_y^{D'}\cdot \int_0^{\Delta z}e^{i(k_z^{D'}-k_z^D)\varepsilon}\, d\varepsilon, \\ (S31)$$

Here, we assume that the scattering potential tensor does not vary within each slice along the axial direction; thus, $\overline{\overline{V}}(x',y',n\Delta z+\varepsilon) \approx \overline{\overline{V}}(x',y',n\Delta z)$. When the thickness of slice $\Delta z$ is small, the integration along slice thickness $\Delta z$ can be approximately replaced by $\Delta z$. The forward birefringent scattering process can be written as:

$$\vec{u}_s^n[k_x^D,k_y^D,(n+1)\Delta z] = -\frac{i}{2}Q(k_x^D,k_y^D,k_z^D)\frac{1}{k_z^D}e^{ik_z^D \Delta z}\Delta z \\ \cdot \iint \overline{\overline{V}}(k_x^D-k_x^{D'},k_y^D-k_y^{D'},n\Delta z)Q(k_x^{D'},k_y^{D'},k_z^{D'})\vec{u}_{in}^n[k_x^{D'},k_y^{D'},n\Delta z]dk_x^{D'}k_y^{D'}, \quad (S32)$$

By using the convolution theorem, the convolution in the frequency domain can be easily calculated in the spatial domain. Here, we rewrite Eq. (S32) by using the vectorial scattering operator **H** and polarization transfer operator **Q**:

$$\vec{u}_s^n[x,y,(n+1)\Delta z] = \mathbf{H}\,\overline{\overline{V}}(x,y,n\Delta z)\mathbf{Q}\,\vec{u}_{in}^n[x,y,n\Delta z], \quad (S33)$$

where the tensor matrix **Q** is the polarization transfer operator:

$$\mathbf{Q} = \mathcal{F}^{-1} Q(k_x,k_y,k_z)\mathcal{F}\{\cdots\}, \quad (S34)$$

and the tensor matrix **H** denotes the vectorial scattering operator:

$$\mathbf{H} = \mathcal{F}^{-1} -\frac{i}{2} Q(k_x,k_y,k_z)\frac{\exp(ik_z\Delta z)}{k_z}\mathcal{F}\{\cdots\}\Delta z, \quad (S35)$$